\documentclass[twocolumn]{aastex62}

\newcommand{\muJy}		{~$\mu$Jy~beam$^{-1}$}
\newcommand{\mJy}		{~mJy~beam$^{-1}$}

\newcommand{\kl}		{~k$\lambda$}
\newcommand{\Msun}		{~$M_{\sun}$}

\newcommand{\ie}    	{i.\,e.,}

\newcommand{\Mm}		{$\pm$}

\graphicspath{{./}}

\shorttitle{A polarized dusty disk around a massive star}
\shortauthors{Girart et al.}

\begin{document}

\title{Resolving the polarized dust emission of the disk around the massive star powering the HH~80-81 radio jet}

\correspondingauthor{J. M. Girart}
\email{girart@ice.cat}

\author[0000-0002-3829-5591]{J. M. Girart}
\affil{Institut de Ci\`encies de l'Espai (ICE, CSIC), Can Magrans s/n, E-08193 Cerdanyola del Vall\`es, Catalonia}
\affiliation{Institut d'Estudis Espacials de de Catalunya (IEEC), E-08034 Barcelona, Catalonia}

\author{M. Fern\'andez-L\'opez}
\affiliation{Instituto Argentino de Radioastronom{\'\i}a, (CCT-La Plata, CONICET; CICPBA), C.C. No. 5, 1894,Villa Elisa, Argentina}

\author{Z.-Y. Li}
\author{H. Yang}
\affiliation{Astronomy Department, University of Virginia, Charlottesville, VA 22904, USA}

\author[0000-0001-7341-8641]{R. Estalella}
\affiliation{Dep. de F\'{\i}sica Qu\`antica i Astrof\'{\i}sica, Institut de Ciències del Cosmos, Universitat de Barcelona, E-08028 Barcelona, Catalonia}
\affiliation{Institut d'Estudis Espacials de de Catalunya (IEEC), E-08034 Barcelona, Catalonia}

\author{G. Anglada}
\affiliation{Instituto de Astrof\'{\i}sica de Andaluc\'{\i}a (IAA, CSIC), Glorieta de la Astronom\'{\i}a s/n, E-18008 Granada, Spain}

\author{N. \'A\~nez-L\'opez}
\affil{Institut de Ci\`encies de l'Espai (ICE, CSIC), Can Magrans s/n, E-08193 Cerdanyola del Vall\`es, Catalonia}

\author[0000-0002-2189-6278]{G. Busquet}
\affil{Institut de Ci\`encies de l'Espai (ICE, CSIC), Can Magrans s/n, E-08193 Cerdanyola del Vall\`es, Catalonia}
\affiliation{Institut d'Estudis Espacials de de Catalunya (IEEC), E-08034 Barcelona, Catalonia}

\author{C. Carrasco-Gonz\'alez}
\affiliation{Instituto de Radioastronom\'{\i}a y Astrof\'{\i}sica (UNAM), 58089 Morelia, M\'exico}

\author{S. Curiel}
\affiliation{Instituto de Astronom\'{\i}a (UNAM), 04510 M\'exico, DF, M\'exico}

\author{R. Galvan-Madrid}
\affiliation{Instituto de Radioastronom\'{\i}a y Astrof\'{\i}sica (UNAM), 58089 Morelia, M\'exico}

\author{J. F. G\'omez}
\affiliation{Instituto de Astrof\'{\i}sica de Andaluc\'{\i}a (IAA, CSIC), Glorieta de la Astronom\'{\i}a s/n, E-18008 Granada, Spain}

\author{I. de Gregorio-Monsalvo}
\affiliation{European Southern Observatory, 3107 Alonso de Cordova, Vitacura, Santiago, Chile}
\affiliation{Joint ALMA Observatory, Alonso de Cordova 3107, Vitacura, Casilla 19001, Santiago 19, Chile}

\author[0000-0003-4493-8714]{I. Jim\'enez-Serra}
\affiliation{Queen Mary University of London, Mile End Road, E1 4NS London, UK}

\author{R. Krasnopolsky}
\affiliation{Institute of Astronomy \& Astrophysics, Academia Sinica, Taipei, Taiwan}

\author{J. Mart\'{\i}}
\affiliation{Departamento de F\'{\i}sica (EPSJ), Universidad de Ja\'en, Campus Las Lagunillas s/n, A3-420, E-23071 Ja\'en, Spain}

\author{M. Osorio}
\affiliation{Instituto de Astrof\'{\i}sica de Andaluc\'{\i}a (IAA, CSIC), Glorieta de la Astronom\'{\i}a s/n, E-18008 Granada, Spain}

\author{M. Padovani}
\affiliation{INAF-Osservatorio Astrofisico di Arcetri, Largo E. Fermi 5, 50125 Firenze, Italy}

\author[0000-0002-1407-7944]{R. Rao}
\affiliation{Institute of Astronomy and Astrophysics, Academia Sinica, 645 N. Aohoku Pl., Hilo, HI 96720, USA}

\author{L. F. Rodr\'{\i}guez}
\affiliation{Instituto de Radioastronom\'{\i}a y Astrof\'{\i}sica (UNAM), 58089 Morelia, M\'exico}

\author{J. M. Torrelles}
\affil{Institut de Ci\`encies de l'Espai (ICE, CSIC), Can Magrans s/n, E-08193 Cerdanyola del Vall\`es, Catalonia}
\affiliation{Institut d'Estudis Espacials de de Catalunya (IEEC), E-08034 Barcelona, Catalonia}


%|textcolor{orange} modifications by Estalella

\begin{abstract}

Here we present deep (16~\muJy), very high (40~mas) angular resolution 1.14 mm, polarimetric, Atacama Large Millimeter/submillimeter Array (ALMA) observations towards the massive protostar driving the HH 80-81 radio jet. The observations clearly resolve the disk oriented perpendicular to the radio jet, with a radius of $\simeq0\farcs171$ ($\sim$291~au at 1.7 kpc distance). The continuum brightness temperature, the intensity profile, and the polarization properties clearly indicate that the disk is optically thick for a radius of $R\la 170$~au. The linear polarization of the dust emission is detected almost all along the disk and its properties suggest that dust polarization is produced mainly by self-scattering. However, the polarization pattern presents a clear differentiation between the inner (optically thick) part of the disk and the outer (optically thin) region of the disk, with a sharp transition that occurs at a radius of $\sim0\farcs1$ ($\sim$170~au). The  polarization characteristics of the inner disk suggest that dust settling has not occurred yet with a maximum dust grain size between 50 and 500~$\mu$m.  The outer part of the disk has a clear azimuthal pattern but with a significantly higher polarization fraction compared to the inner disk. This pattern is broadly consistent with self-scattering of a radiation field that is beamed radially outward, as expected in the optically thin outer region, although contribution from non-spherical grains aligned with respect to the radiative flux cannot be excluded. 

\end{abstract}


\section{Introduction} \label{sec:intro}

HH~80-81 is a spectacular, 14~pc long and highly collimated radio jet  powered  by a massive (early B-type) protostar  \citep[D$\simeq$1.7~kpc][]{Rodriguez1980, Marti93, Marti95, Masque13, Masque15, Vig18}.  The protostellar radio jet is the first one where linearly polarized synchrotron emission due to relativistic electrons has been detected, indicating the presence of a magnetic field  aligned with the jet \citep{Carrasco10, RodriguezK17}.  This is highly indicative that the jet is being launched from an accretion disk, as it has been observed in other astrophysical environments (e.g., microquasars and AGNs).  Previous pre-ALMA sub-arcsecond angular resolution observations reveal the presence of a $\sim$1000~au rotating molecular flattened structure perpendicular to the radio jet \citep{Gomez03, FernandezG11, Carrasco12, Girart17} and of a compact, barely resolved dust emission from the putative disk \citep{FernandezC11, Carrasco12, Girart17}. The dynamical mass (star and disk) derived from these observations is roughly 10--20\Msun\  \citep[e.g.,][]{Girart17}.  These observations put an upper limit for the dust linear polarization of $\simeq 0.8$\% at circumstellar scales. However,  the dust polarization is detected at much larger  scales \citep[$\sim$0.1~ pc,][]{Curran2007}.

Circumstellar disk polarization studies at (sub)mm wavelengths have undergone a great development since the early, pre-ALMA, interferometric observations \citep{Rao2014, Stephens2014, SeguraCox2015, Cox2015, Fernandez16}. Now ALMA is revealing in great detail the rich and complex polarization information hidden in the dust particles at circumstellar disk scales, where grain growth, dust settling and optical depth effects may have a significant impact on the observed polarization properties \citep[e.g.,][]{HullMocz17, HullGirart17, Stephens2017, Lee18, Cox18, Maury2018}. This is leading to new challenges on how we understand the production of (sub)mm polarization from a theoretical perspective. Indeed, three main mechanisms are used currently to explain the dust polarization disk at (sub)mm wavelengths: self-scattering by dust grains \citep{Kataoka2015, Kataoka2016, Yang2016a, Yang17}, alignment of aspherical dust grains with the magnetic field \citep{Lazarian2008, Hoang2008}, and alignment with anisotropic radiation \citep{LazarianHoang2007, Zeng2013, Tazaki2017}. However, the dust polarization in disks could arise from a combination of them \citep{Yang2016b, Kataoka2017}, which makes the interpretation difficult.
% * <jm.girart@gmail.com> 2018-02-21T17:52:31.322Z:
% 
% I have deleted the Phol2017 since they present near-IR pol, where scattering is produced by small grains
% 
% ^.

Here we present 1.14~mm polarimetric ALMA  observations of the accretion disk of the massive protostar GGD27~MM1 powering the  HH~80-81 jet (Section 2). The disk has been fully resolved using a beam of $\sim$40~mas (Section 3). Continuum polarization observations are presented as well, showing in detail the complex morphology of the polarized emission (Section 3). We discuss the implication of these findings in Section 4. 

\section{Observations}

The 1.14 mm (263.0 GHz) ALMA continuum polarimetric observations were taken on December 3, 2015. We used the Band 6 receiver with the correlator set in continuum mode (time division mode, TDM) covering the 253.0--257.0~GHz and 269.0--273.0~GHz frequency ranges. The observations were performed with 37 antennas in the C36-7 configuration which provided baselines between 15 m and 6.3 km (or 13 to 5400\kl). The observations were performed under very good  weather conditions, which yielded system temperatures between 80 and 100~K at 263~GHz. The standard aperture synthesis calibration and the specific polarization calibration were performed by the ALMA staff \citep[see][]{Nagai2016}. The ALMA flux accuracy in Band 6 is $\sim$10\%, as determined by the observatory flux monitoring program. 

The observations have a very good coverage in the visibility plane except in the 150 and 300\kl\ range, where there is a significant lack of visibilities. That means that the image fidelity is severely affected for scales larger than $\ga 0\farcs7$. Because of this, iterative phase-only selfcalibration was performed using the Stokes I image as a model for the visibilities with baselines larger than 300\kl. The  selfcalibration solutions were transfer to the Stokes Q and U visibilities. The selfcalibration allowed to improve the rms noise in Stokes map (by a factor of $\sim$2 for the Stokes I). 

In this paper we present the polarization maps towards GGD27 MM1, all obtained with the CASA task {\it clean}.  The  Stokes I, Q and U maps were generated with a value of 0.5 for the robust Briggs weighting parameter, and only using visibilities with baselines larger than 300\kl. The resulting synthesized beam has a full width at half maximum (FWHM) of 45.0 mas$\times$38.3 mas with a position angle of $-62.4^\circ$.  The Stokes I rms noise is 19\muJy. The rms noise level in the Stokes Q and U dust maps is $\sigma_{\rm pol}=$16~\muJy. The linear polarization maps were obtained from the Stokes Q and U images ($P = \sqrt{Q^2 +U^2 -\sigma_{\rm pol}^2}$). The polarization position angles ($PA_{\rm pol} = 0.5 \arctan{\rm (U/Q)}$) were obtained using a cutoff of 3-$\sigma$, where $\sigma$ is the rms noise of the Stokes Q and U maps.

\section{Results}

\subsection{Dust emission}\label{Sec:Dust}

\begin{figure}[htbp]
\begin{center}
\includegraphics[width=0.9\columnwidth]{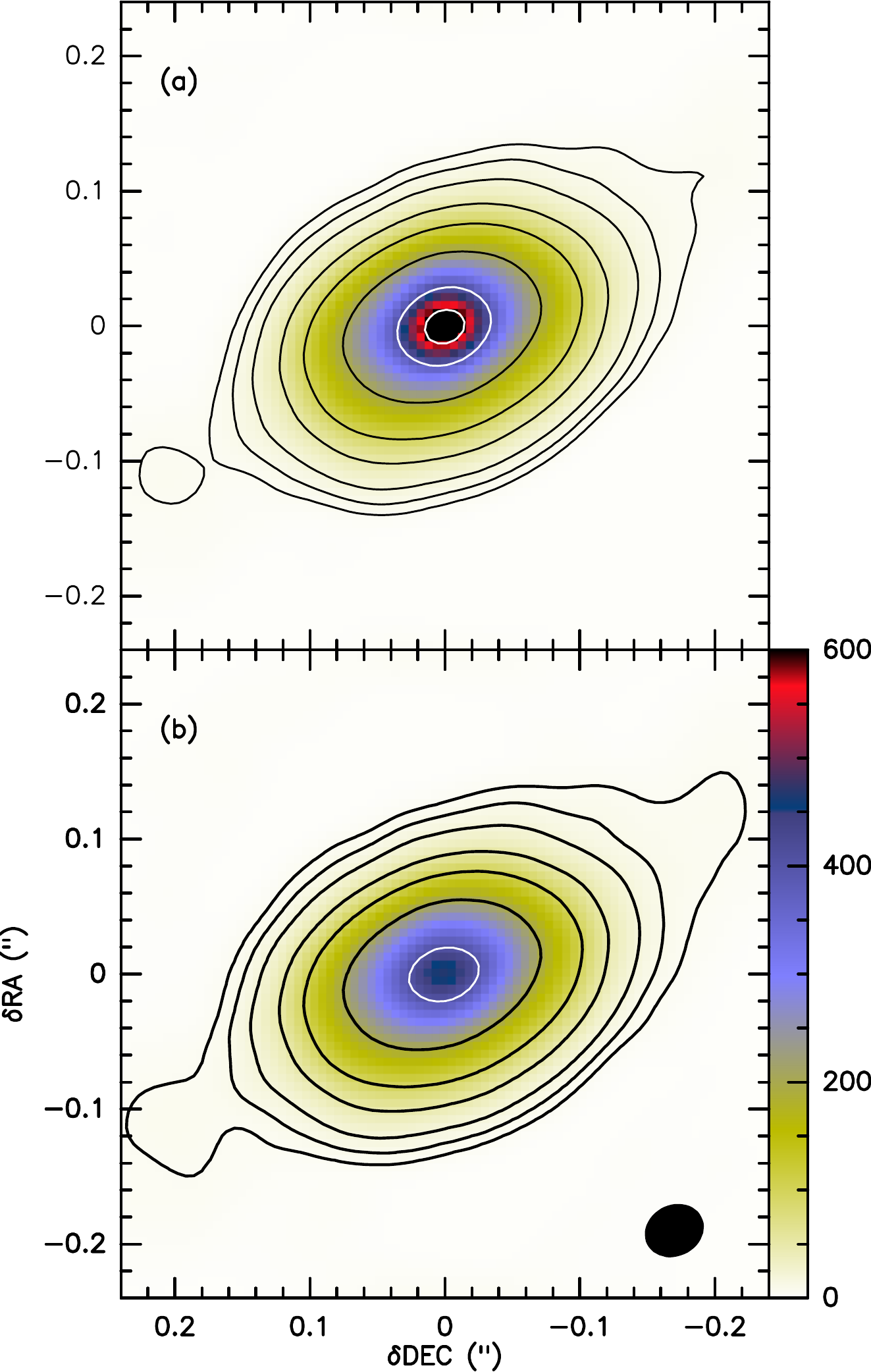}\\
\caption{
Image of the GGD 27 MM1 disk in units of brightness temperature. (a) Original map; (b) Map obtained after subtracting the compact source from the visibilities. The color scale is the same for the two panels. The black contours are 5, 10, 20, 50, 100, and 200 K, the white contours 400 and 600~K. The synthesized beam is shown in the right-bottom corner of the bottom panel.}
\label{fig:MM1}
\end{center}
\end{figure}

The 1.14~mm continuum dust emission of GGD27 MM1 appears clearly resolved in an elliptical shape of $\sim$$0\farcs4\times0\farcs2$ elongated along the NW-SE direction, perpendicular to the radio jet (Figure~\ref{fig:MM1}).  The source radius, $\sim$300~au, suggests that the emission is tracing the expected putative disk around the massive protostar that powers the HH~80-81 jet.  The peak intensity is 65.9\mJy,  which implies a signal-to-noise ratio of $\sim$3500. The brightness temperature at the peak for the achieved angular resolution is 676~K.

\begin{figure}[htbp]
\begin{center}
\includegraphics[width=1.0\columnwidth]{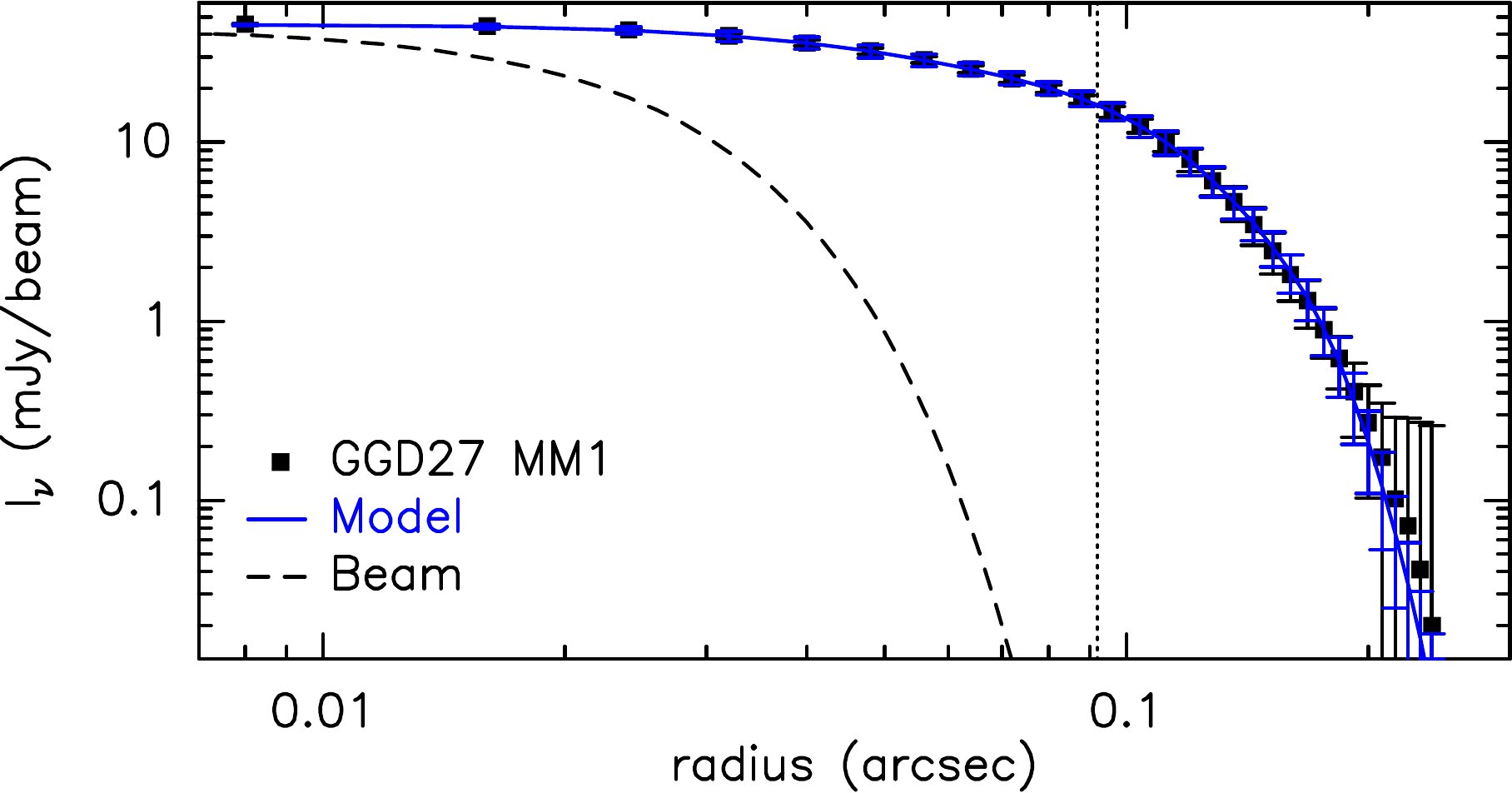}\\
\caption{
Radial profile of the 1.14 mm dust emission from the GGD27 MM1 disk, obtained by averaging the emission over elliptical rings (filled squares). The blue and dashed line show the best fitted profile (see Section~\ref{Sec:Dust}) and the beam profile, respectively. The error bars show the rms dispersion within each ring for the data. The vertical dotted line indicates $R_{\rm turn}$, the radius where the power-law changes.
}
\label{Fig:Profile}
\end{center}
\end{figure}

% Gauss fit towards MM1 disk (without compact src): Bad residual map
% Position: $12\fs09554$ & $30\farcs9475$ &
% Peak/flux 41.82\Mm0.10 & 356.6\Mm0.97 &
% Deconvolved size 137.5\Mm0.4 $\times$92.6\Mm0.3		& 112.6\Mm0.3  \\

We found that in the maps obtained using only baselines larger than $\ga 4000$\kl\ the emission appears to arise from an unresolved, unpolarized source. This means that the source size is much smaller than the synthesized beam and that it has a very high brightness temperature, $\gg$750~K. The origin of this emission, probably ionized gas associated with the base of the jet, will be analysed in a forthcoming publication (A\~nez et al., in preparation). Therefore, the compact source was removed from the visibilities and the maps were redone with the same parameters (hereafter, we use maps without the compact source). Figure~\ref{fig:MM1} shows the disk before and after the removal. The overall shape of the disk remains the same, but the peak brightness temperature decreases from 676~K to 467~K.   

%Paragraph revised by Estalella 2017/02/24
Since the disk is very well resolved, a Gaussian cannot reproduce the intensity 2-D profile. Alternatively, we assumed that the disk intensity can be modeled with two different power laws ($I_{\nu} \propto r^{-q}$)\footnote{A fit using a simple power law generates a significantly worse solution.}, one for the inner region (with a power law index $q_{\rm inn}$) and the other for the outer region ($q_{\rm out}$). We define three radii: $R_{\rm inn}$ is the minimum radius where the dust emission is detected, $R_{\rm turn}$, the radius where the power-law changes, and $R_{\rm disk}$, the disk radius. As free parameters we also included the disk position angle, $PA$, and the angle between the disk normal and the line-of-sight, $i$. The disk model was convolved with a Gaussian with the same FWHM as the synthesized beam. The best fit was obtained for a disk with a radius of 171\Mm6~mas (291\Mm10~au), $PA=112.9$\Mm1.0\degr, $i=48.8$\Mm0.5\degr\ and a total 1.14 mm flux density of 350\Mm1~mJy (see Fig.~\ref{Fig:Profile}).  The inner and transition  radius are $R_{\rm inn}=7$\Mm1~mas (12$\pm$2~au) and $R_{\rm turn}=92$\Mm1 mas (156~au), respectively.  In addition, the power-law index changes dramatically in the disk, from a steep index in the outer part ($q_{\rm out}$=4.5\Mm0.1) to a relative flat index in the inner disk ($q_{\rm inn}$=0.7\Mm0.1). Previous 7~mm observations marginally resolved the dust emission associated with this disk, but the derived radius, $\sim 200$~au, had large uncertainty due to the lack of sensitivity and to the  contamination from the radio jet emission \citep{Carrasco12}. The 1.14 and 7~mm flux densities indicate a spectral index of 2.2, which implies that the disk dust emission is nearly optically thick.

\subsection{Dust polarization}\label{SecDustPol}

\begin{figure}[htbp]
\begin{center}
\includegraphics[width=1.0\columnwidth]{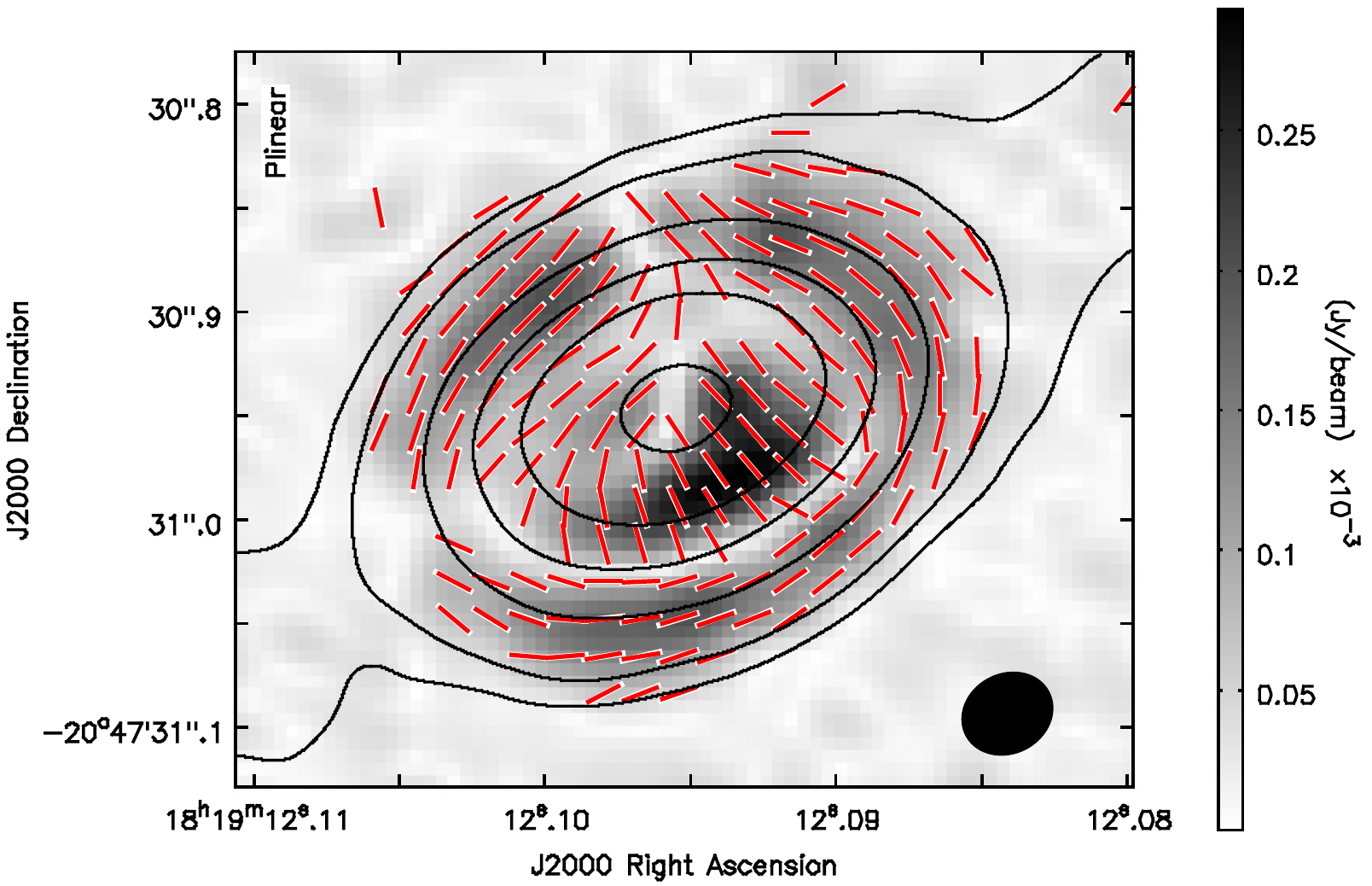}\\
\includegraphics[width=1.0\columnwidth]{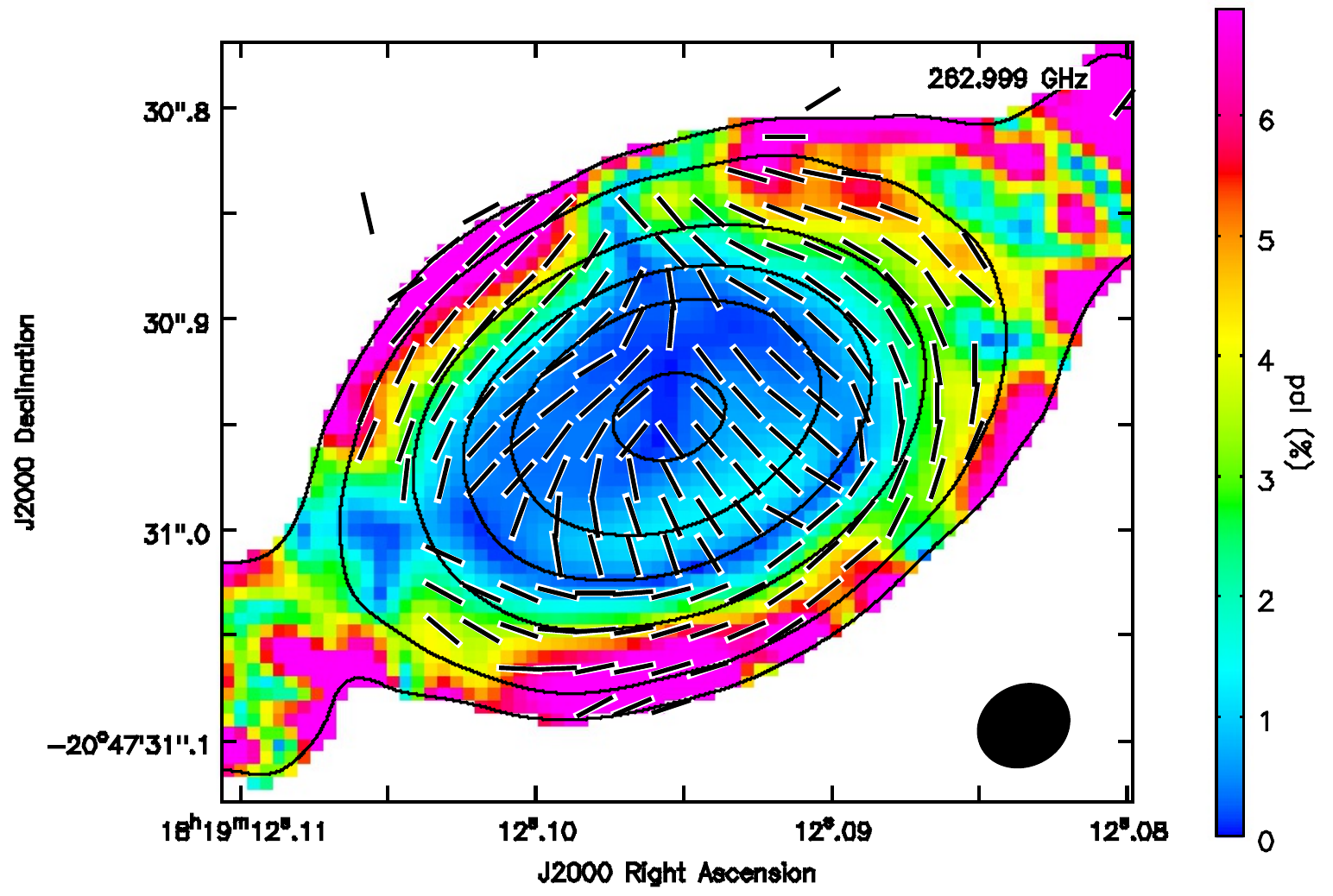}\\
\caption{
{\it Top panel: } Composite of the 1.14~mm continuum ALMA Stokes I (contours), polarized intensity (grey scale) and polarization position angle segments (red bars) in the GGD27 MM1 disk. {\it Bottom panel: } Similar as before, but the colour image showing the polarization fraction (here, the polarization position angle segments  appear as black bars).  In both panels, the contour levels are 3, 10, 50, 120, 200 and 400~K. The wedge shows the scale of the polarized intensity and of polarization fraction. The synthesized beam is shown in the bottom right corner of the panels. Note that the 120~K contour appears at the transition radius, with an almost null polarization ring, between two distinctive polarization patterns.
}
\label{fig:MM1pol}
\end{center}
\end{figure}

% 3213.8,3205.4  - 3219.9, 3200.8 = 6.1, 4.6  R = 7.64pix 
%
% Null of emission between outer/inner occurs at 15% contour level of Stokes I peak intensity.
% Mean = 0.57% stdev = 0.36%, Median = 0.35%
The dust linear polarization is detected almost all along the GGD27 MM1 disk (Fig.~\ref{fig:MM1pol}). The polarized peak intensity, 0.28\mJy\ ($\simeq$0.67\% of the Stokes I peak intensity), is offset 45~mas south-west of the Stokes I intensity peak. Remarkably, the polarization properties show two clear distinct regions in the disk with a sharp transition between them, which coincides with a location of zero-level polarization. This occurs at the Stokes I iso-intensity contour level of 11.8\mJy, or, equivalently, at a dust brightness temperature of 120~K.  Interestingly, this transition is located at a disk radius of  $R\simeq0\farcs10$, which is precisely the transition radius, $R_{\rm turn}$, where the power-law of the Stokes I intensity profile changes (see previous section). 

\begin{figure}[htbp]
\begin{center}
\includegraphics[width=\columnwidth]{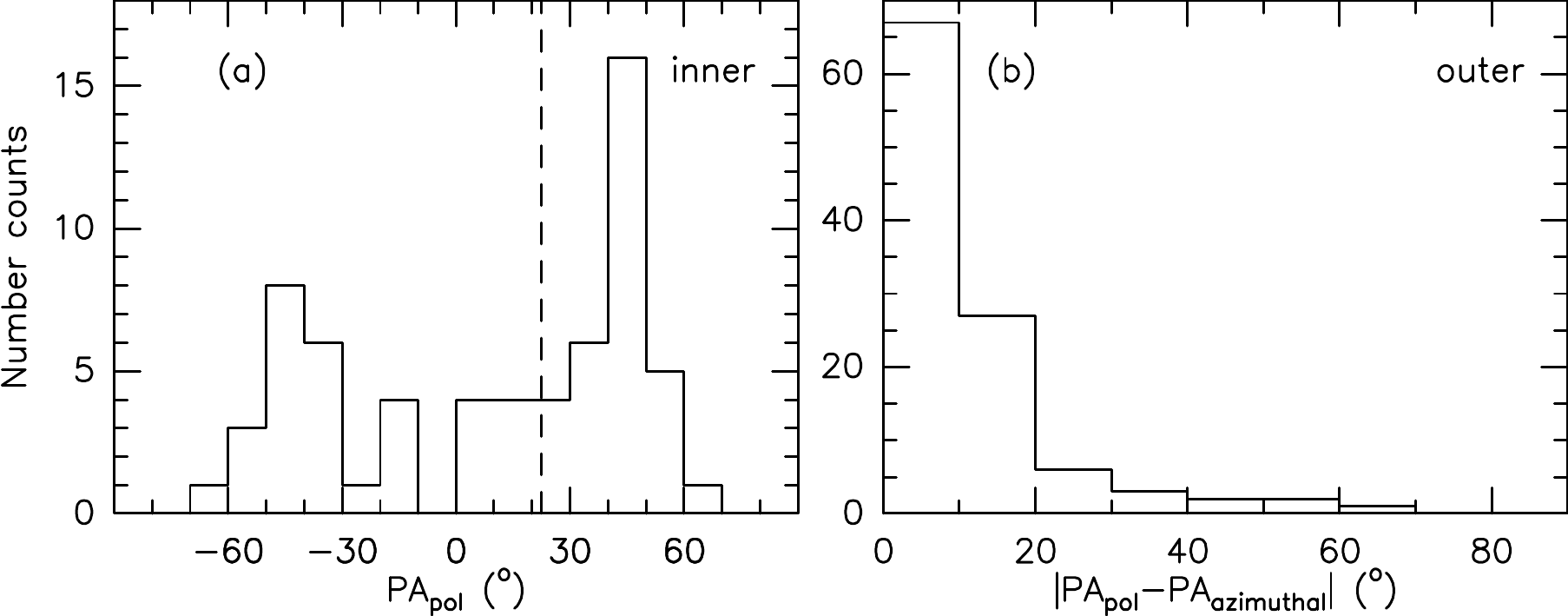}\\
\caption{
{\it Left panel:} Histogram of the distribution of the polarization position angles for the disk's inner region ($R<0\farcs10$). The dashed line indicates the position angle of the disk minor axis.
{\it Right panel:}  Histogram of the differences between the polarization position angles and the disk azimuthal angles for the disk's outer region ($R>0\farcs10$). 
}
\label{fig:PAdistrib}
\end{center}
\end{figure}

The region inside the transition radius, $R_{\rm turn}$, (hereafter the inner disk) has a very low polarization level region. Thus, within this radius the polarization degree has a mean value of 0.56\% and a standard deviation of 0.33\%. In the left panel of Figure~\ref{fig:PAdistrib}, the histogram of the polarization position angles shows two main directions. Most of the angles are partially aligned with the minor axis (22.6\degr), with an average direction of  $\sim36$\degr. The other region arises from the north-east quadrant of the inner disk, and shows an average position angle of $-38$\degr.  

The outer disk (i.e., the region outside $R_{\rm turn}$) has significantly higher polarization levels than the inner disk, with an average value of 4.0\% and standard deviation of 2.3\% \footnote{This has been measured in a region enclosed between $R_{\rm turn}$ and the disk radius of $R_{\rm disk}$.}.  The polarization position angles have a nearly azimuthal pattern (i.e., they follow a direction tangent to the iso-intensity contours of the Stokes I emission). At the position of each polarization position angle segment shown in Fig.~\ref{fig:MM1pol}, we have computed the expected tangent angle for an ellipse centered in the disk center that crosses this point. The right panel of Figure~\ref{fig:PAdistrib} shows the histogram of the difference between this tangent angle and the observed polarization angle. The average angle difference is relatively small, 11\degr. This confirms the azimuthal pattern in the outer disk.

\section{Discussion}

\subsection{The optically thick disk around the massive protostar powering the HH~80-81 jet}\label{Ddisk}

The brightness temperature of the disk dust emission reaches high values, up to $\sim400$~K, but it drops well below 10~K in the outskirts of the disk. This suggests that the inner part of the disk is optically thick and that the dust optical depth decreases with radius, becoming optically thin.   A multi line analysis of SO$_2$, at scales of a few hundred au, indicates that the average kinetic temperature of the traced molecular gas is $\sim120$~K \citep{FernandezG11}. Thus, we can roughly use this temperature as a proxy to define where the dust emission becomes optically thick, i.e., where the dust brightness temperature reaches a value similar to the gas  temperature (assuming that gas and dust are well coupled). The dust brightness temperature reaches this value at the $R_{\rm turn}$ radius, \ie\ at the radius where the Stokes intensity profile changes significantly in its slope, so it becomes much flatter in the inner disk with respect to the outer disk.  In addition, this is the radius where there is a sharp transition between the polarization properties. Together, all these pieces of evidence strongly suggest that the $R_{\rm turn}$ radius ($0\farcs1$ or 170~au) indicates the change between the optically thick and the optically thin regimes in the disk. This implies that observations at 1.14~mm or at shorter wavelength cannot be used to properly study the disk density and temperature as well as the kinematics in the inner disk.  Observations at longer wavelengths may alleviate somewhat this problem.

\subsection{Origin of the dust polarization}\label{ScDpol}

Because of the clearly different polarization properties of the inner and outer regions of the disk, here we discuss separately the properties of the two regions.  

\begin{figure}[htbp]
\begin{center}
\includegraphics[width=0.9\columnwidth]{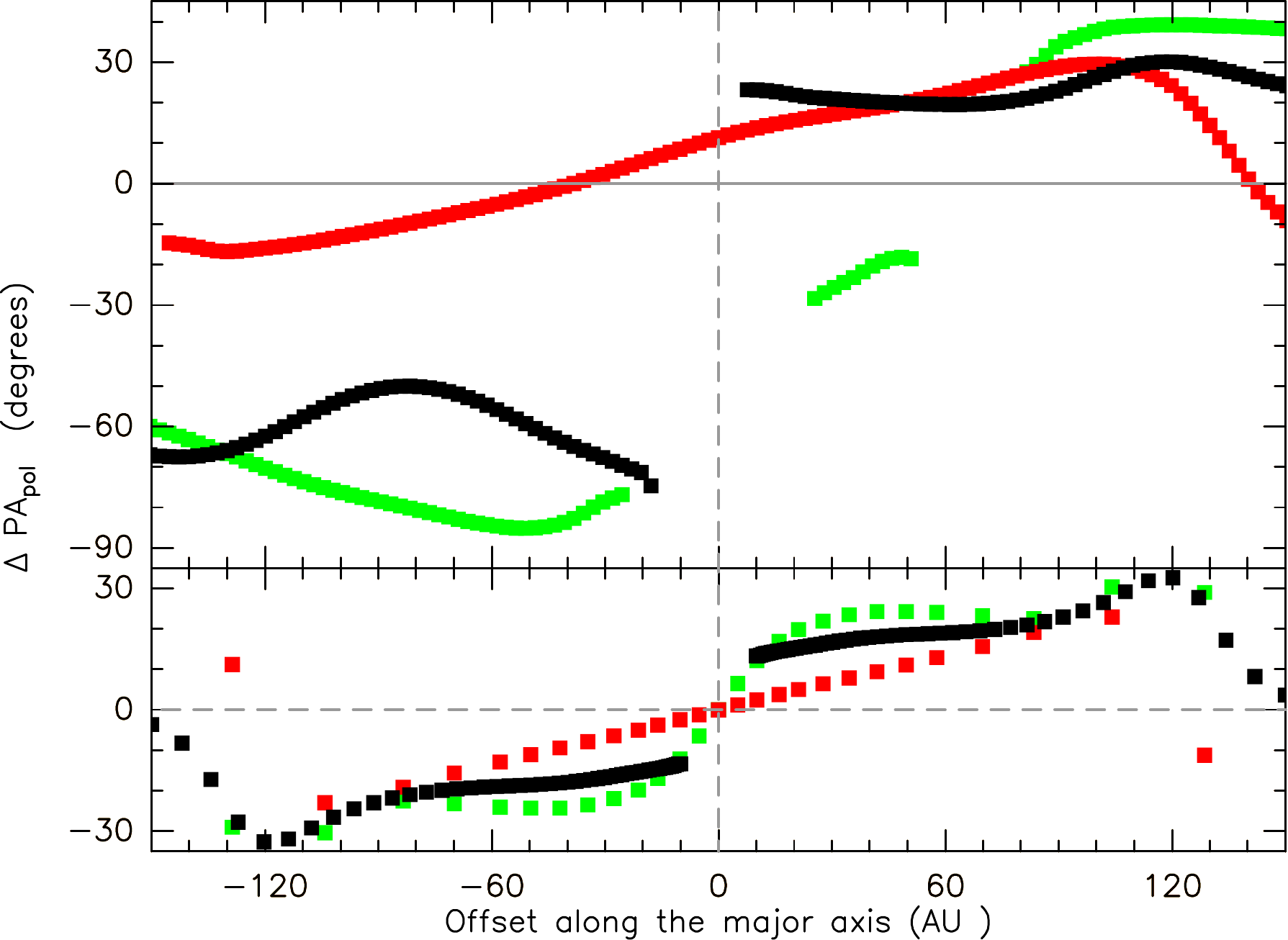}\\
\caption{
{\it Top:}  Difference between the position angles of the dust polarization and the disk's minor axis angle for three cuts parallel to the major axis ($PA=113$\degr): one along the major axis itself (black line), a cut 40~mas (68~au) north of the major axis (in the far side of the disk, green line), and a cut 40~mas south of the major axis (in the near side of the disk, red line).  Positive values of the position offset indicates position west of the disk intensity peak.
{\it Bottom:}  Similar but from the optically thick disk model by \citet{Yang17}.
}
\label{fig:PA_MajAx}
\end{center}
\end{figure}

The overall polarization pattern of the inner disk at first sight matches well the prediction of self-scattering from an optically thick disk where dust settling (into the mid-plane) has not yet occurred \citep{Yang17}. First, the most clear sign of the optically thick case is that the strongest polarization signal is offset along the minor axis with respect to the total intensity peak. Not only this, the strongest polarized emission is extended along the major axis with a curved shape, similar to what is expected \citep[see Fig.~4h from][]{Yang17}. In addition, the outflow geometry \citep[e.g.,][]{Heathcote98} indicates that  this region  appears in the nearest side of the disk (with respect to the observer), which is also a strong prediction by the aforementioned model. Another strong indication of self-scattering in an optically thick disk is the bifurcation in polarization orientation, with respect to the minor axis direction, along the major axis. Thus, in an optically thin disk the orientation is basically uniform and parallel to the minor axis. In the optically thick case, where only the surface layer of a (dust) disk of an appreciable geometric thickness is directly  observable, the polarization orientation is predicted to deviate significantly from that of the minor axis in a well defined manner.  The observed position angles in the inner disk show a reasonable agreement with this predicted pattern, except on its NE side (see Section~\ref{SecDustPol}). This can be better observed by making slices along the major axis and along a parallel line to the major axis in the near and far sides of the disk (see Fig.~\ref{fig:PA_MajAx}). For the case of an optically thick disk, polarization angles in cuts parallel to the major axis are expected to be symmetrical with respect to the origin (defined as the peak position relative to the minor axis), as it is shown in the bottom panel of Fig.~\ref{fig:PA_MajAx}. This behavior is partially observed in the data. In spite of being slightly offset, the slice in the near side of the disk is the one that matches better the expected shape. The other two slices show that the position angles in the eastern side deviate significantly, by 30 to 50\degr, with respect to the expected values. One possibility for this significant departure from the predicted scattering pattern is that the (dust) disk surface is highly perturbed, potentially by a disk-wind or some other means. Another possibility is that they are modified by polarized emission from, e.g., magnetically aligned grains, in the envelope surrounding the disk.

The average polarization fraction detected in the inner region, 0.56\%, is well within the expected values for self-scattering, but it is lower than the predicted average maximum values \citep{Kataoka2016, Yang17}. A rough estimate of the maximum grain size in the GGD27 MM1 disk can be obtained from the studies carried out by \citet[and references therein]{Kataoka2016}. From Figure 3 of \citet{Kataoka2016} and taking into account the average polarization fraction in GGD27 MM1, we estimate that the maximum grain size should be in the 50 and 500~$\mu$m range. 

In the outer part of the disk, the azimuthal pattern is remarkable (Figs.~\ref{fig:MM1pol} and \ref{fig:PAdistrib}b). This is expected for the outer layers of the disk, where the radially anisotropic radiation is basically dominated by the radiation from the inner part of the disk. In models of optically thin disks  around low mass stars, this layer is typically very thin with a  polarization degree higher than in the center, $\sim$2--3 \% \citep{Kataoka2016,Yang17}. The GGD27 MM1 disk presents three clear differences with respect to these published predictions. First, the azimuthal layer is significantly broader and appears to extend outwards starting from the transition radius between optically thick and thin regimes. Second, the polarization fraction reaches in many zones values of 5-7$\%$, which appears to be higher than the predicted values. Third, the polarization fraction in the outer part of the disk is apparently larger along the minor axis than along the major axis. It is possible that these discrepancies with the existing scattering models (that are not tailored for our particular source) can be resolved with an outer disk where the temperature drops more quickly with radius than assumed in the existing models. We have carried out some preliminary models that confirm this hypothesis (a more detailed study is out of the scope of this paper and will be part of a future work). It is also plausible that the polarization in the outer disk is due to non-spherical grains aligned by anisotropic radiation \citep{Tazaki2017}. 

Finally, this work has shown the unique case of a fully linearly polarized disk around a massive protostar.  Observations at multiple wavelengths but with similar angular resolution are needed to better constrain the polarization origin in the outer disk, the maximum grain size in the inner and outer disk, and to determine the origin of  the SE-NW asymmetry in the inner disk.

%% NO SUMMARY SECTION. 

\acknowledgments

This paper makes use of the following ALMA data: ADS/JAO.ALMA\#2015.1.00480.S. ALMA is a partnership of ESO (representing its member states), NSF (USA) and NINS (Japan), together with NRC (Canada) and NSC and ASIAA (Taiwan) and KASI (Republic of Korea), in cooperation with the Republic of Chile. The Joint ALMA Observatory is operated by ESO, AUI/NRAO and NAOJ.
JMG, RE, GA, IA, GB, IA, JFG, MO and JMT are supported by the MINECO (Spain)  AYA2014-57369-C3 and AYA2017-84390-C2 coordinated grants. SC acknowledges support from DGAPA, UNAM and CONACyT, M\'exico. HY is supported in part by ALMA SOS, and ZLY by NASA  NNX14AB38G and NSF AST-1313083 and  1715259. MP acknowledges funding from the European Unions Horizon 2020 research and innovation programme under the Marie Sk\l{}odowska-Curie grant agreement No 664931. IJS acknowledges financial support from STFC through an Ernest Rutherford Fellowship (ST/L004801). JM acknowledges support from MINECO (Spain) AYA2016-76012-C3-3-P grant.

\vspace{5mm}
\facilities{ALMA}
\software{CASA \citep{CASA07}}

%\bibliography{girart}



\end{document}